\documentstyle[epsf,12pt,fleqn]{article}

\def\Slash#1{{\rm\ooalign{\hfil$/$\hfil\crcr \hbox{$#1$}}}}
\newcommand\Kmbf[1]{\mbox{\boldmath{$#1$}}}

\begin{document}

\begin{flushright}
 TIT/HEP-425/NP
\end{flushright}

\begin{center}
{\Large \bf
$U_A(1)$ Symmetry Breaking and $\eta,\eta'$ mesons 
\hfil\break
in the Bethe--Salpeter Approach} 
\par
\par
\par
\vskip 5mm
K. Naito\footnote{E-mail address: kenichi@th.phys.titech.ac.jp}\\
{\it Radiation Laboratory, the Institute of Physical and Chemical 
Research (Riken),}\\
{\it Wako, Saitama 351-0198, Japan} \\
Y. Nemoto\\
{\it Research Center for Nuclear Physics (RCNP), Osaka University,}\\
{\it Ibaraki, Osaka 567-0047, Japan}\\
M. Takizawa\\
{\it Laboratory of Computer Sciences, 
Showa College of Pharmaceutical Sciences,} \\
{\it Machida, Tokyo 194-8543, Japan}\\
K. Yoshida and M. Oka\\
{\it Department of Physics, Tokyo Institute of Technology,} \\
{\it Meguro, Tokyo 152-8551, Japan} 
\end{center}
\vskip 5mm
\begin{abstract}
\baselineskip 1.5pc
 $U_A(1)$ symmetry breaking is studied by introducing the 
flavor mixing interaction proposed by 
Kobayashi, Maskawa and 't Hooft.
Combining the one gluon exchange interaction, the rainbow like
Schwinger--Dyson equation and the ladder like Bethe--Salpeter 
equation are derived.
The anomalous PCAC relation in the framework of this 
approximation is considered. 
The masses of the pseudoscalar
mesons $\pi,\eta$ and $\eta'$ are calculated.
It is found that the pion mass is not sensitive to the strength of 
the flavor mixing interaction.  On the other hand, the masses of $\eta$ 
and $\eta'$ are reproduced by a relatively weak flavor mixing interaction, 
for which the chiral symmetry breaking is dominantly induced by the
soft-gluon exchange interaction.   The decay constants are calculated
and the anomalous PCAC relation is numerically checked.  It is found
that the flavor structures of the $\eta$ and $\eta'$ mesons
significantly depend on their masses and therefore it is quetionable to
define a flavor mixing angle for $\eta$ and $\eta'$. 
\end{abstract}
\section{Introduction} \label{SEC:H100926:1}
It is known that the classical QCD lagrangian is invariant under the 
$U_L(3) \times U_R(3)$ symmetry except for the quark mass term,
and this symmetry is broken down to the $U_V(3)$ spontaneously 
in the low-energy QCD. In this case, the number of the Nambu-Goldstone 
bosons should be nine.
However, the number of the observed light pseudoscalar mesons is eight.
The ninth pseudoscalar meson, $\eta'$ meson, is heavier than the other octet 
pseudoscalar mesons ($\pi^0$, $\pi^+$, $\pi^-$, $K^+$, $K^-$, $K^0$, 
$\overline{K}^0$ and $\eta$) which are well identified 
with the Nambu--Goldstone bosons.  This is the well-known $U_A(1)$ 
problem \cite{W75}.
It is solved by realizing that the $U_A(1)$ symmetry is broken by 
the anomaly.  The phenomena related to the $U_A(1)$ anomaly in the 
low-energy QCD have been studied in the following approaches.
\par
   The first one is the $1/N_C$ expansion approach \cite{tH74,C84}.  
The key point is that the effect of the $U_A(1)$ anomaly is higher order in 
the $1/N_C$ expansion and the low-energy effective lagrangian of QCD has
been derived \cite{VV80}.  Recently, the expansion in powers of 
$1/N_C$, momenta and quark masses was extended to the first non-leading 
order \cite{L96} and the reasonable description of the nonet pseudoscalar
mesons was obtained. 
\par
    The second is the instanton approach \cite{tH86}. 
The instanton is a classical solution of the Euclidean
Yang--Mills equation and may contribute a large weight in the 
Feynman path integration. In the presence of the light quarks, instantons 
are associated with fermionic zero modes which give rise to the $U_A(1)$ 
symmetry breaking. In the dilute instanton gas approximation, the $U_A(1)$ 
breaking 6-quark flavor determinant interaction was derived in the three 
flavor case \cite{tH76}. 
This approach has been developed to the instanton liquid 
picture of the QCD vacuum \cite{SS98}.  In this picture, the instanton 
plays a crucial role in understanding not only the $U_A(1)$ anomaly but also
the spontaneous breaking of the chiral symmetry itself.
\par
    In the third approach, the effective low-energy quark models of QCD
were used to study the structure of the hadrons.  
The introduction of the instanton induced six-quark interaction to the 
effective quark model is one of the handy way to incorporate the 
the $U_A(1)$ breaking effects in the low-energy effective quark model 
of QCD.  
The Nambu-Jona-Lasinio (NJL) model \cite{NJL} is one of the 
simplest and widely used model in studying the structure of the
Nambu-Goldstone bosons.  Using the three-flavor NJL model with the 
instanton induced six quark interaction, properties of the nonet 
pseudoscalar mesons were investigated \cite{KH88}.  
A shortcoming of this approach is that the $\eta'$ mass has unphysical 
imaginary part associated with the unphysical decay channel 
$\eta' \rightarrow \bar qq$.
Recent study of the radiative decays of the $\eta$ meson in the 
NJL model has shown that the observed values of the mass and 
radiative decay amplitudes are reproduced well with a rather 
strong $U_A(1)$ breaking interaction \cite{TO95}. 
Such strong $U_A(1)$ breaking may be consistent with the instanton liquid 
picture of the QCD vacuum.
\par
    In contrast with the instanton liquid model, the study of the 
QCD Schwinger-Dyson (SD) equation for the quark propagator in the 
improved ladder approximation (ILA) has shown that the spontaneous
breaking of the chiral symmetry is explained by simply extrapolating 
the running coupling constant from the perturbative high-energy region to 
the low-energy region \cite{HM84}. 
\par
    Then, the Bethe-Salpeter (BS) equation for the $J^{PC} = 0^{-+}$ 
$q \bar q$ channel has been solved in the ILA and the existence of 
the Nambu-Goldstone pion has been confirmed \cite{KG1}.
The numerical predictions of the pion decay constant $f_{\pi}$ and the
quark condensate $\langle \overline{\psi} \psi \rangle$ are rather good.
It has been also shown that the BS amplitude shows the correct
asymptotic behavior as predicted by the operator product expansion (OPE) 
in QCD \cite{KG2}.  
The masses and decay constants for the lowest lying scalar, vector and 
axial-vector mesons have been evaluated by calculating
the two point correlation functions for the composite operators
$\overline{\psi} M \psi$.  The obtained values are in good agreement with
the observed ones \cite{KG4}.  
\par
   Recently, the current quark mass term has 
been introduced in the studies of the BS amplitudes in the ILA \cite{NYNOT3}
and the reasonable values of the pion mass, the pion decay constant and 
the quark condensate have been obtained with a rather large 
$\Lambda_{\rm QCD}$. 
It has been also shown that the pion mass square and the pion decay 
constant are almost proportional to the current quark mass up to 
the strange quark mass region.
\par
    Since the $\eta$ and $\eta'$ system is expected to be sensitive 
to the $U_A(1)$ anomaly, the study of the $\eta$ and $\eta'$ structure
may give us information on the roles of the $U_A(1)$ anomaly in the
low-energy QCD.  The purpose of this paper is to study the properties 
of the $\eta$ and $\eta'$ mesons by solving the coupled channel 
BS equation in the ILA.  The effect of the $U_A(1)$ anomaly is introduced 
by the instanton induced six-quark determinant interaction.  The instanton 
size effects are taken into account by the form factor of the interaction 
vertices. It guarantees the right asymptotic behavior of the solutions of the 
SD and BS equations.
\par
    There have been many studies of the pion BS amplitude using the 
effective models of QCD and /or the approximation schemes of QCD 
\cite{RW94}. As for the $\eta$ and $\eta'$ system, Jain and Munczek 
model \cite{JM93} has been applied to them \cite{KK98}. 
They have introduced the effect of the $U_A(1)$ anomaly 
by simply adding the additional mass term in the flavor singlet 
pseudoscalar meson channel by hand and the reasonable values of the 
masses and decay constants have been obtained. 
\par
    It is known that the introduction of the two gluon exchange 
diagrams in the calculation of the $\eta$ and $\eta'$ BS amplitudes
beyond the ladder approximation
do not break the $U_A(1)$ symmetry with the perturbative gluon propagator.
Recently, it has been shown that if the gluon propagator has the strong 
infrared singularity, the $U_A(1)$ symmetry breaks \cite{FM98}.
The relaton between this approach and the instanton approach is not 
clear.
\par
    The paper is organized as follows.  In Sec.~\ref{SEC:H110107:1} we 
explain the model Lagrangian we have used in the present study. 
In Sec.~\ref{SEC:EA} the Conwall-Jackiw-Tomboulis (CJT) effective action 
\cite{CJT74} calculated from our model Lagrangian is presented. 
The SD and BS equations are derived from the CJT effective action in 
Sec.~\ref{SEC:SD} and Sec.~\ref{SEC:BS}.  
In Sec.~\ref{SEC:DC} the meson decay constant is derived.
In Sec.~\ref{SEC:NGS} the Nambu-Goldstone solution of the BS equation 
is derived and the anomalous PCAC relation is discussed in 
Sec.~\ref{SEC:PCAC}. Sec.~\ref{SEC:NR} is devoted to the numerical results.  
Finally, summary and concluding remarks are given in Sec.\ref{SEC:SC}.
\section{Model} \label{SEC:H110107:1}
 We use the flavor three $(N_F=3)$ effective model whose 
lagrangian density is given by
\begin{equation}
 {\cal L}[\psi,\overline{\psi}] 
 := \overline{\psi} f(\partial^2) (i\Slash{\partial}-m_0)\psi
 + {\cal L}_{\rm GE}[\psi,\overline{\psi}] 
 + {\cal L}_{\rm FM}[\psi,\overline{\psi}],
 \label{AEQ:H110107:1}
\end{equation}
\begin{equation}
 \psi :=(u,d,s)^T \label{AEQ:H110107:2}
\end{equation}
where $f(\xi)$ is a cut-off function defined by \cite{NYNOT3}
\begin{equation}
 f(\xi) = 1 + M \theta(\xi-\Lambda_{\rm UV}^2),\quad M\to \infty.
 \label{AEQ:H100319:11}
\end{equation}
$m_0$ denotes a diagonal quark mass matrix $m_0={\rm diag}(m_q,m_q,m_s)$
under the assumption of the isospin invariance.
${\cal L}_{\rm GE}$ denotes a 
gluon exchange interaction 
\begin{eqnarray}
   {\cal L}_{\rm GE}[\psi,\overline{\psi}](x) &:=& 
   -\frac{1}{2} \int_{pp'qq'} {\cal K}^{mm',nn'}(p,p';q,q') \nonumber \\
   & & {} \times \overline{\psi}_m(p)\psi_{m'}(p') 
   \overline{\psi}_n(q) \psi_{n'}(q') e^{-i(p+p'+q+q')x} \, ,
      \label{AEQ:H110315:5}
\end{eqnarray}
\begin{eqnarray}
   {\cal K}^{mm',nn'}(p,p';q,q') &=& \bar{g}^2 
   \left( (\frac{p_E-q_E'}{2})^2,(\frac{q_E-p_E'}{2})^2\right )  \nonumber \\
   &&\times iD^{\mu\nu} \left(\frac{p+p'}{2}-\frac{q+q'}{2} \right)
   (\gamma_\mu T^a)^{mm'} (\gamma_\nu T^a)^{nn'} 
      \label{AEQ:H100315:6}
\end{eqnarray}
where $\int_p$ denotes $\int \frac{d^4p}{(2\pi)^4}$ and $p_E$ represents
the Euclidean momentum. The indeces $m, n, \dots$ represent combined 
indeces in the color, flavor and Dirac spaces.
In Eq.(\ref{AEQ:H100315:6}) we employ the Landau gauge gluon propagator, 
\begin{equation}
 iD^{\mu\nu}(k)  = \left( g^{\mu\nu} - \frac{k^\mu k^\nu}{k^2} \right)
 \frac{-1}{k^2} ,
 \label{AEQ:H100319:10a}
\end{equation}
and the Higashijima--Miransky type  
running coupling constant $\bar g^2$  
defined as follows.
\begin{equation}
   \bar{g}^2(p_E^2,q_E^2) = \theta(p_E^2-q_E^2) g^2(p_E^2) +
   \theta(q_E^2-p_E^2) g^2(q_E^2),
   \label{AEQ:H100319:7}
\end{equation}
with
\begin{equation}
 g^2(p_E^2) := \left\{
   \begin{array}{ll} \displaystyle{\frac{1}{\beta_0}\frac{1}{1+t}} &
   \mbox{ for } t_{\rm IF}\le t \\
   \mbox{} & \\
   \displaystyle{\frac{1}{2\beta_0}\frac{1}{(1+t_{\rm IF})^2}
   \left[3t_{\rm IF}-t_0+2-\frac{(t-t_0)^2}{t_{\rm IF}-t_0}\right]} &
   \mbox{ for } t_0\le t \le t_{\rm IF} \\
   \mbox{} & \\
   \displaystyle{\frac{1}{2\beta_0}
   \frac{3t_{\rm IF}-t_0+2}{(1+t_{\rm IF})^2}}&
   \mbox{ for } t\le t_0 \end{array} \right. ,
   \label{AEQ:H100319:8}
\end{equation}
\begin{equation}
 t := \ln \frac{p_E^2}{\Lambda_{\rm QCD}^2} - 1 ,
 \label{AEQ:H100319:9}
\end{equation}
\begin{equation}
 \beta_0 := \frac{1}{(4\pi)^2}\frac{11N_C-2N_f}{3}. 
 \label{AEQ:H100319:10}
\end{equation}
In Eq.(\ref{AEQ:H100319:8}) the infrared cut-off $t_{\rm IF}$ is introduced.
Above $t_{\rm IF}$, $g^2(p_E^2)$ develops according to the one-loop
result of the QCD renormalization group equation and below 
$t_0$, $g^2(p_E^2)$ is kept constant. 
These two regions are connected by the quadratic polynomial so that
$g^2(p_E^2)$ becomes a smooth function.  
Here $N_C$ is the number of colors and $N_f$ is the number of active flavors. 
We use $N_C = N_f = 3$ in our numerical studies.
\par
  ${\cal L}_{\rm FM}$ is the $U_A(1)$ symmetry breaking flavor mixing 
interaction (FMI), our interest, given by
\begin{eqnarray}
 {\cal L}_{\rm FM}[\psi,\overline{\psi}](x) & = & 
 \frac{1}{3} G_D \epsilon^{f_1f_2f_3} \epsilon^{g_1g_2g_3}
 \bar w \left( \partial_{x_1}; \partial_{y_1}; \partial_{x_2}; 
 \partial_{y_2}; \partial_{x_3}; \partial_{y_3} \right)
 \nonumber \\
 & & \times \biggm\{ [ \overline{\psi}_{g_1}(x_1)
 \psi_{f_1}(y_1) ][ \overline{\psi}_{g_2}(x_2) \psi_{f_2}(y_2) ]
 [ \overline{\psi}_{g_3}(x_3) \psi_{f_3}(y_3) ] \nonumber \\
 & & {} + 3 [ \overline{\psi}_{g_1}(x_1) \psi_{f_1}(y_1) ]
 [ \overline{\psi}_{g_2}(x_2) \gamma_5 \psi_{f_2}(y_2) ]
 [ \overline{\psi}_{g_3}(x_3) \gamma_5 \psi_{f_3}(y_3) ] 
 \biggm\} \biggm|_*   
\label{AEQ:H110107:3}
\end{eqnarray}
where $f_1,g_1,\cdots$ are flavor indices, $\epsilon$ denotes
the antisymmetric tensor with $\epsilon^{uds}=1$ and the symbol $*$ 
at the end of the equation means
$x_1,y_1,\cdots \to x$ after all derivatives are operated. 
This type of the $U_A(1)$ symmetry breaking six-quark interaction has
been introduced in Ref. \cite{KM70} before the discovery of the 
instanton induced interacion \cite{tH76}.  There is a minor difference 
of the flavor-spin structure between ${\cal L}_{\rm FM}$ and 't Hooft 
instanton induced interaction. In the rainbow-ladder approximation, 
these two interactions give the same effects on the $\eta$-$\eta'$ system.
\par
  We introduce a weight function $\bar w(\cdots)$ which is 
necessary so that FMI is turned off  
at the high energy. We use the following separable Gaussian form
\begin{equation}
 \bar w( \partial_{x_1};\partial_{y_1}
 ;\partial_{x_2};\partial_{y_2};\partial_{x_3};\partial_{y_3}
 )=
 w\left( \frac{\partial_{x_1}^2+\partial_{y_1}^2+
 \partial_{x_2}^2+\partial_{y_2}^2+\partial_{x_3}^2+\partial^2_{y_3}
 }{2} \right),
 \label{AEQ:H110126:1}
\end{equation} 
\begin{equation}
 w(\mu^2) := \exp( -\kappa \mu^2 ).
 \label{AEQ:H110107:4}
\end{equation}
This weight function is convenient for a numerical calculation 
as it satisfies the association rule
\begin{equation}
 w(-p^2-q^2-k^2) = w(-p^2)w(-q^2)w(-k^2). \label{AEQ:H110107:5}
\end{equation}
But this particular set of the momenta in the argument of the 
weight function
modifies the form of the Noether current for the axial-vector 
transformation.
This is the same problem occurred in the Higashijima--Miransky 
approximation in the ${\cal L}_{\rm GE}$ term discussed
extensively in Ref.\cite{NYNOT3}. 
The explicit form of the Noether axial-vector current in this model 
is very complicated and we do not show it.
We treat the exact Noether axial-vector current within 
the ladder like approximation. We will show that the (modified) 
Ward--Takahashi identity for axial-vector current and 
the PCAC relation holds. 
This approach is studied in Ref.\cite{NYNOT1}.
\par
    On the other hand, if one does not want to modify
the Noether current, one has to 
employ an appropriate form of the argument of the 
running coupling constant, such as 
\begin{equation}
 \bar g^2(\cdots) = \bar g^2\left( \left(\frac{p_E+p_E'}{2}-
 \frac{q_E+q_E'}{2} \right)^2 \right)
 \label{AEQ:H110323:2}
\end{equation}
in ${\cal L}_{\rm GE}$ and similarly
\begin{equation}
 \bar w(\cdots) = w\left( \left( \frac{\partial_{x_1}+
 \partial_{x_2}+\partial_{x_3} - \partial_{y_1} - \partial_{y_2} 
 - \partial_{y_3} }{2\sqrt{3}} \right)^2 \right)
 \label{AEQ:H100126:2}
\end{equation}
in ${\cal L}_{\rm FM}$. 
\par
 In our model, there are nine axial-vector currents, 
$J_{5\mu}^\alpha(\alpha=0,\cdots,8)$, which satisfy 
the anomalous PCAC relation
\begin{equation}
 \partial^\mu J^\alpha_{5\mu}(x) = 
 2 [m_0 J_5]^\alpha(x) + \delta^{\alpha 0} A(x),
 \label{AEQ:H100126:3}
\end{equation}
\begin{equation}
 [m_0 J_5]^\alpha := \overline{\psi} i\gamma_5 
 \frac{ f(\stackrel{\leftarrow}{\partial}{\!\!}^2) m_0 \lambda^\alpha
  + \lambda^\alpha m_0 f(\partial^2) }{4}
  \psi \label{AEQ:H100126:4}
\end{equation}
where $\lambda^\alpha$ denotes the Gell-Mann matrix in the
flavor space. $A(x)$ corresponds to the explicit $U_A(1)$ symmetry
breaking and is proportional to 
$G_D$ in Eq.$(\ref{AEQ:H110107:3})$.
In QCD, $A(x)$ is given by 
\begin{equation}
 \frac{3\alpha_S}{8\pi} \epsilon^{\mu\nu\rho\sigma}
 F^{a}_{\mu\nu} F^a_{\rho\sigma}.
 \label{AEQ:H110323:1}
\end{equation}
\section{Effective Action}\label{SEC:EA}
 To derive the Schwinger--Dyson equation and the Bethe--Salpeter
equation, we use the Cornwall--Jackiw--Tomboulis (CJT) effective
action formulation \cite{CJT74}.
In Ref.\cite{NYNOT3}, we have already derived the CJT effective action
in the lowest order (rainbow--ladder) approximation in the framework
of the ILA model. Here we add a new term $\Gamma_{\rm FM}[S_F]$ 
which contains the lowest order effect of the flavor mixing 
interaction (FMI).
\begin{equation}
 \Gamma[S_F] := i{\rm Tr}{\rm Ln}[S_F] -i{\rm Tr}[S_0^{-1} S_F]
  + \Gamma_{\rm GE}[S_F] + \Gamma_{\rm FM}[S_F]
  \label{AEQ:H110201:1}
\end{equation}
 $\Gamma_{\rm GE}[S_F]$ corresponds to the two-loop (eyeglass) diagram using 
gluon exchange interaction and is defined by 
Eq.$(15)$ in Ref.\cite{NYNOT3}.
 Since FMI is a six-quark interaction, the lowest two particle
irreducible vacuum diagram of FMI is a three loop (clover) diagram.
For simplicity, we take only the dominant term in $1/N_C$ 
expansion for $\Gamma_{\rm FM}[S_F]$ term as
\begin{eqnarray}
 \Gamma_{\rm FM}[S_F] & = & \frac{G_D}{3}\int d^4 x 
 \epsilon^{f_1f_2f_3}\epsilon^{g_1g_2g_3}w\left(\frac{
 \partial^2_{x_1}+\partial^2_{y_1}+\partial^2_{x_2}
 +\partial^2_{y_2}+\partial^2_{x_3}+\partial^2_{y_3}}{2}\right)
 \label{AEQ:H091213:15} \\
 & & \times \bigg\{ -{\rm tr}^{\rm (DC)}[S_{Ff_1g_1}(y_1,x_1)]
 {\rm tr}^{\rm (DC)}[S_{Ff_2g_2}(y_2,x_2)]
 {\rm tr}^{\rm (DC)}[S_{Ff_3g_3}(y_3,x_3)] \nonumber \\
 && {}-3{\rm tr}^{\rm (DC)}[S_{Ff_1g_1}(y_1,x_1)]
 {\rm tr}^{\rm (DC)}[\gamma_5 S_{Ff_2g_2}(y_2,x_2)]{\rm tr}^{\rm (DC)}
 [\gamma_5 S_{Ff_3g_3}(y_3,x_3)] \bigg\} \Bigg|_* . \nonumber 
\end{eqnarray}
 In this approximation, the global $SU_L(3) \times SU_R(3)$
symmetry is preserved. 
In fact, the total effective action is invariant under the
infinitesimal global chiral transformation
\begin{eqnarray}
 S_F(x,y) & \to & (1+i\frac{\lambda^\alpha}{2}\theta^\alpha ) S_F(x,y)
 (1-i\frac{\lambda^\alpha}{2}\theta^\alpha ), \label{AEQ:H110201:2} \\
 S_F(x,y) & \to & (1+i\gamma_5\frac{\lambda^\alpha}{2}\theta^\alpha) 
 S_F(x,y) (1+i\gamma_5\frac{\lambda^\alpha}{2}\theta^\alpha) 
 \label{AEQ:H110201:3}
\end{eqnarray}
except for the quark bare mass term and the flavor mixing term which breaks the $U_A(1)$ symmetry.
\section{Schwinger--Dyson Equation}\label{SEC:SD}
\begin{figure}[tbp]
  \centerline{ \epsfxsize=12cm \epsfbox{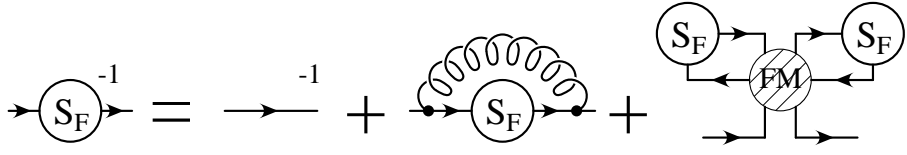} }
  \caption{The diagram for the SD equation.}
  \label{FIG:H110209:1}
\end{figure}

 The Schwinger--Dyson equation is derived by the stability condition 
of the CJT effective action
\begin{equation}
 \frac{\delta \Gamma[S_F]}{\delta S_{Fmn}(x,y)} = 0.
 \label{AEQ:H110202:1}
\end{equation}
The detailed procedure is same as in Ref.\cite{NYNOT3}.
Introducing the regularized propagators 
$S_F^R(q) := f(q^2) S_F(q)$ and $S_0^R(q) := i/(\Slash{q}-m_0)$,
the SD equation in momentum space becomes
\begin{eqnarray}
\lefteqn{ i{S^R_F}^{-1}_{nm}(q) - i{S^R_0}^{-1}_{nm}(q) = - 
        \frac{C_F}{f(-q^2)} \int_p \frac{1}{f(-p^2)}
 \bar{g}^2(q_E^2,p_E^2) iD^{\mu\nu}(p-q) 
 ( \gamma_\mu S^R_{Fm_2n_2}(p)\gamma_\nu )_{nm} } \nonumber \\
 & & {}+ G_D \delta_{ij}\delta_{ab} 
 \epsilon^{gf_1f_2}\epsilon^{fg_1g_2}
 \int_{p,k}\frac{1}{f(-p^2)f(-k^2)} w(-q^2-p^2-k^2) \nonumber \\
 & & {} \times {\rm tr}^{\rm (DC)}[S^R_{Fg_1f_1}(p)]
 {\rm tr}^{\rm (DC)}[S^R_{Fg_2f_2}(k)] \label{AEQ:H091214:2}
\end{eqnarray}
where the indices $m,n,\cdots$
are combined indices $m:=(a,i,f),\,n:=(b,j,g),\cdots$ with 
Dirac indices $a,b,\cdots$ and color indices $i,j,\cdots $ 
and flavor indices $f,g,\cdots$.
This equation is shown diagramatically in Fig.\ref{FIG:H110209:1}.
Generally the quark propagator is parametrised by
\begin{equation}
 S^R_{Fh}(q) = \frac{i}{\Slash{q}A_h(q^2) - B_h(q^2)}
 \label{AEQ:H110202:2}
\end{equation}
where the index $h$ denotes the flavor.
 After the Wick rotation, we obtain $A_h(-q_E^2)\equiv 1$.
Then the resulting SD equation reads
\begin{eqnarray}
 B_q(-s) & = & m_q + \frac{3C_F}{16\pi^2}\int_{0}^{\Lambda_{\rm UV}^2}dr 
 \bar{g}^2(s,r)\frac{B_q(-r)}{r+B_q^2(-r)}
 \left\{ \theta(r-s)+\frac{r}{s}\theta(s-r)\right\}  \nonumber \\
 & & {} -\frac{G_D N_C^2 w(s)}{8\pi^4} \int_{0}^{\Lambda_{\rm UV}^2} 
 dr w(r) \frac{r B_q(-r)}{r+B_q^2(-r)} \int_{0}^{\Lambda_{\rm UV}^2}
  dl w(l) \frac{l B_s(-l)}{l+B_s^2(-l)}, \label{AEQ:H091216:11} 
\end{eqnarray}
\begin{eqnarray}
 B_s(-s) & = & m_s + \frac{3C_F}{16\pi^2}\int_{0}^{\Lambda_{\rm UV}^2}dr 
 \bar{g}^2(s,r)\frac{B_s(-r)}{r+B_s^2(-r)} 
 \left\{ \theta(r-s)+\frac{r}{s}\theta(s-r)\right\}  \nonumber \\
 & & {} -\frac{G_D N_C^2 w(s)}{8\pi^4} \int_{0}^{\Lambda_{\rm UV}^2}
  dr w(r) \frac{r B_q(-r)}{r+B_q^2(-r)} \int_{0}^{\Lambda_{\rm UV}^2}
  dl w(l) \frac{l B_q(-l)}{l+B_q^2(-l)} , 
  \label{AEQ:H091216:12}
\end{eqnarray}
with $s \equiv q_E^2$.
These integral equations are solvable numerically.
\par
    Since the improved ladder approximation (ILA) model reproduces the 
asymptotic behavior of QCD, the quark mass function can be renormalized so 
that the solution of the SD equation is matched with the QCD quark mass 
function in the aymptotic region. We renormalize the quark mass function 
properly in the manner described in Ref.\cite{NYNOT3}.
The renormalization constants $Z_{m_q}$ and $Z_{m_s}$ defined by 
$m_q = Z^{-1}_{m_q}m_{qR}$ and $m_s = Z^{-1}_{m_s}m_{sR}$ are determind 
by the condition
\begin{equation}
 \frac{\partial B_q(\mu^2)}{ \partial m_{qR} }\bigg|_{m_{qR}=0} = 1
 \quad \mbox{and}
 \quad 
 \frac{\partial B_s(\mu^2)}{ \partial m_{sR} }\bigg|_{m_{sR}=0} = 1.
\end{equation} 
Note that the flavor mixing interaction (FMI) does not disturb the 
asymptotic behavior of the ILA model and QCD because of the Gaussian type 
weight function.
\section{Bethe--Salpeter Equation}\label{SEC:BS}
\begin{figure}[tbp]
  \centerline{ \epsfxsize=12cm \epsfbox{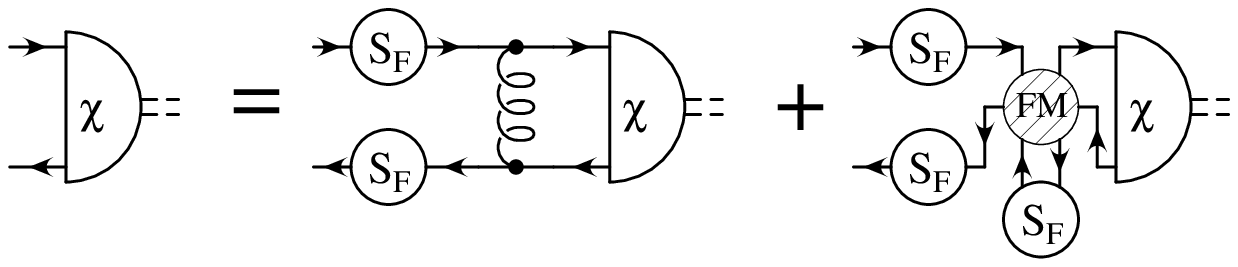} }
  \caption{The diagram for the BS equation.}
  \label{FIG:H110209:2}
\end{figure}

 The homogeneous Bethe--Salpeter (BS) equation is derived by 
\begin{equation}
 \frac{ \delta^2 \Gamma[S_F] }{\delta S_{Fmn}(x,y) \delta S_{Fn'm'}(y',x')}
 \chi_{n'm'}(y',x';P_B) = 0
 \label{AEQ:H110209:1}
\end{equation}
where 
\begin{equation}
 \chi_{n'm'}(y',x';P_B) = \langle 0 | T \psi_{n'}(y') 
 \overline{\psi}_{m'}(x') | \Kmbf{P} \rangle 
 \label{AEQ:H110209:2}
\end{equation}
denotes the BS amplitude. 
The normalization condition is 
$\langle \Kmbf{P}_B |  \Kmbf{P}'_B \rangle = 
(2\pi)^3 2P_{B0}\delta^3(\Kmbf{P}_B - \Kmbf{P}'_B)$ and
$P_B:=(\sqrt{M_B^2+\Kmbf{P}_B^2},\Kmbf{P}_B)$ is the on-shell
momentum of the meson.  
 Introducing the regularized BS amplitude by
\begin{equation}
 \chi^R_{nm}(q;P_B) := f(-q_+^2) \chi_{nm}(q;P_B) f(-q_-^2),
  \label{AEQ:H110209:4}
\end{equation}
\begin{equation}
  q_+ := q+\frac{P_B}{2} ,\quad q_- := q - \frac{P_B}{2},
  \label{AEQ:H110209:8}
\end{equation}
the BS equation in momentum space becomes 
\begin{eqnarray}
 \lefteqn{  {S^R_F}^{-1}_{nn_1}(q_+) \chi^R_{n_1m_1}(q;P_B)
 {S^R_F}^{-1}_{m_1m}(q_-) } \nonumber \\
 & = & -iC_F \int_k  \frac{1}{f(-k_+^2) f(-k_-^2)}
  \bar{g}^2(q_E^2,k_E^2) iD^{\mu\nu}(q-k)  ( \gamma_\mu
  \chi^R(k;P_B) \gamma_\nu )_{nm} \nonumber \\
 & & {} +2i G_{\rm D} \epsilon^{ghf'} \epsilon^{fh'g'} \delta_{ji} 
  \int_{p,k} \frac{1}{f(-p^2)f(-k_+^2) f(-k_-^2)}
w\left(-p^2-q^2-k^2 - \frac{ P^2_B}{2} \right) \nonumber  \\
 & & \times   {\rm tr}^{\rm (DC)}[S^R_{Fh'h}(p)] \Big\{(\gamma_5)_{ba}
 {\rm tr}^{\rm (DC)}[\gamma_5 \chi^R_{g'f'}(k;P_B)] 
 + 1_{ba} {\rm tr}^{\rm (DC)}[\chi^R_{g'f'}(k;P_B)] \Big\}
 \label{AEQ:H110209:3}
\end{eqnarray}
which is shown diagramatically in Fig.\ref{FIG:H110209:2}.
For the pseudoscalar state $|\Kmbf{P}_B\rangle$, the last term 
in the braces does not contribute.
\par
    For the pion, the BS amplitude can be written in terms of 
four scalar amplitudes as in Ref.\cite{NYNOT3},
\begin{eqnarray}
 \lefteqn{ 
 \chi^R_{nm}(k;P) = \delta_{ji} \frac{(\lambda^{\alpha})_{gf}}{2}\bigg[
  \bigg(\phi_S(k;P) + \phi_P(k;P) \Slash{k} + \phi_Q(k;P)\Slash{P} 
  } \nonumber \\
  & & \quad {}  +\frac{1}{2}\phi_T(k;P)(\Slash{P}\Slash{k}-
  \Slash{k}\Slash{P})\bigg)\gamma_5\bigg]_{ba}
 \label{AEQ:H110209:5}
\end{eqnarray}
where $\lambda^\alpha$ denotes the flavor structure of the pion state.
For example, the neutral pion is given by $\alpha=3$
\begin{equation}
 \lambda^3 = \pmatrix{ 1 & 0 & 0 \cr 0 & -1 & 0 \cr 0 & 0 & 0}.
 \label{AEQ:H110301:1}
\end{equation} 
On the other hand, for the $\eta$ and $\eta'$ mesons,
 the BS amplitudes are written in 
terms of eight scalar amplitudes,
\begin{eqnarray}
 \lefteqn{ 
 \chi^R_{nm}(k;P) = \delta_{ji} \frac{(\lambda^{qq})_{gf}}{2}\bigg[
  \bigg(\phi^{qq}_S(k;P) + \phi^{qq}_P(k;P) \Slash{k} 
 + \phi^{qq}_Q(k;P)\Slash{P} 
  } \nonumber \\
  & & \quad {}  +\frac{1}{2}\phi^{qq}_T(k;P)(\Slash{P}\Slash{k}-
  \Slash{k}\Slash{P})\bigg)\gamma_5\bigg]_{ba} \\
 & & {} + \delta_{ji} \frac{(\lambda^{ss})_{gf}}{2}\bigg[
  \bigg(\phi^{ss}_S(k;P) + \phi^{ss}_P(k;P) \Slash{k} 
  + \phi^{ss}_Q(k;P) \Slash{P}    \nonumber \\
  & & \quad {}  +\frac{1}{2}\phi^{ss}_T(k;P)(\Slash{P}\Slash{k}-
  \Slash{k}\Slash{P})\bigg)\gamma_5\bigg]_{ba}
 \label{AEQ:H110209:6}
\end{eqnarray}
where the flavor matrices $\lambda^{qq},\lambda^{ss}$ are defined by
\begin{equation}
 \lambda^{qq} = \pmatrix{ 1 & 0 & 0 \cr 0 & 1 & 0 \cr 0 & 0 & 0 } ,\quad
 \lambda^{ss} = \pmatrix{ 0 & 0 & 0 \cr 0 & 0 & 0 \cr 0 & 0 & \sqrt{2} }. 
 \label{AEQ:H110209:7}
\end{equation}
This is because the $qq$ and $ss$ components of the BS amplitude 
are mixed through the last term
of the right hand side in Eq.$(\ref{AEQ:H110209:3})$ which is FMI.
The BS equations for the $\eta$ and $\eta'$ mesons are same.
The ground state solution is identified with the $\eta$ meson and the first
excited state solution is identified with the $\eta'$ meson.
Substituting Eq.$(\ref{AEQ:H110209:5})$ or Eq.$(\ref{AEQ:H110209:6})$ into
Eq.$(\ref{AEQ:H110209:3})$, we obtain the coupled integral equations.
The explicit form is rather complicated and we do not show it. 
Formally the equations can be written as 
\begin{equation}
 \phi_{\cal A} (q;P_B) = \int_k M_{\cal A B}(q,k;P_B) \phi_B(k;P_B).
 \label{AEQ:H110209:9}
\end{equation}
$\phi_{A}$ or $\phi_{B}$ denotes the set of amplitudes,
$\phi_S,\phi_P,\phi_Q,\phi_T$ for the pion and 
$\phi^{qq}_S,\phi^{qq}_P,\phi^{qq}_Q,\phi^{qq}_T$,
$\phi^{ss}_S,\phi^{ss}_P,\phi^{ss}_Q,\phi^{ss}_T$ for the $\eta$ 
and $\eta'$ meson. Instead of solving Eq.$(\ref{AEQ:H110209:9})$ directly,
we solve an eigenvalue problem 
\begin{equation}
 \lambda \phi_{\cal A} (q;P_B) = \int_k M_{\cal A B}(q,k;P_B) \phi_B(k;P_B)
 \label{AEQ:H110301:2}
\end{equation}
for a fixed $P_B^2 = M_B^2 = -P_{BE}^2 = -M_E^2 <0$. Then we plot and
extrapolate the eigenvalue $\lambda$ as a function of $P_B^2$
and search for the on-shell point $\lambda=1$.
\section{Decay Constant}\label{SEC:DC}
 To obtain the decay constant, we need the normalization 
of the BS amplitude which is derived from the inhomogeneous 
BS equation. In Ref.\cite{NYNOT3} the normalization condition 
in the momentum space is given by
\begin{equation}
 \lim_{P\to P_B} 
 i \int_{q\in I} \chi^R_{n_1m_1}(q;P_B)
\overline{\chi}^R_{m_2n_2}(q;P_B) 
\frac{P^\mu}{-2P^2} \frac{\partial}{\partial P^\mu}\left[
 S^{R-1}_{Fn_2n_1}(q+\frac{P}{2}) S^{R-1}_{Fm_1m_2}(q-\frac{P}{2}) 
 \right] = 1 
 \label{AEQ:H110210:5}
\end{equation}
where the integral region is 
\begin{equation}
 I = \{ q \,|\, {-(q \pm P_B)^2} \le \Lambda_{\rm UV}^2 \}.
 \label{AEQ:H110210:6}
\end{equation}
Using the normalized BS amplitude, the decay constant is obtained by
\begin{eqnarray}
 \lefteqn{ f_B = \lim_{P\to P_B} \frac{1}{P^2}
 \int_q \frac{1}{f(-q^2_-)f(-q^2_+)}
 {\rm tr}\bigg[ \overline{\chi}^R(q;P_B)
 i\gamma_5\frac{\lambda^\alpha}{2}\bigg\{
 \frac{ f(-q^2_-) + f(-q_+^2) }{2} \Slash{P} } \nonumber \\
 & & + (-f(-q^2_-)+f(-q^2_+) )\Slash{q}
 \bigg\} \bigg] \nonumber \\
& & {} + \lim_{P\to P_B} \frac{1}{P^2} \int_q
 \frac{1}{f(-q^2_-) f(-q^2_+)}
 \left\{ {\rm tr}[\overline{\chi}^R(q;P) 
 (E^\alpha(q;P)+F^\alpha(q;P))] \right\}
 \label{AEQ:H110315:2}
\end{eqnarray}
where
\begin{eqnarray}
 E_{mn}^\alpha(q;P) & := &
 \int_k \frac{1}{f(-k^2)} \left[ \quad
  \left\{ {\cal K}^{n'n,mm'}
  \left(-k,q-\frac{P}{2};-q-\frac{P}{2},k+P \right) 
  \right. \right. \nonumber \\
 & & \quad\quad\quad
 - \left. {\cal K}^{n'n,mm'}\left(-k,q-\frac{P}{2};-q+\frac{P}{2},k \right)
  \right\}
 \left(i\gamma_5\frac{\lambda^\alpha}{2}S^R_F(k) \right)_{m'n'} \nonumber \\
 & & \quad\quad
 + \left\{ {\cal K}^{n'n,mm'}
   \left(-k+P,q-\frac{P}{2};-q-\frac{P}{2},k \right) \right. \nonumber \\
 & & \quad\quad\quad
 - \left. \left. {\cal K}^{n'n,mm'}
   \left(-k,q+\frac{P}{2};-q-\frac{P}{2},k \right) \right\}
   \left(S_F^R(k) i \gamma_5 \frac{\lambda^\alpha}{2}\right)_{m'n'} \right]
 \label{AEQ:H110315:3} \\
 F_{mn}^\alpha(q;P) & := & 
 2G_D (\gamma_5)_{ab}\delta_{ij} \int_{k,l} \frac{1}{f(-k^2)f(-l^2)}
   {\rm tr}^{\rm (DC)}[S^R_{Fg_2f_2}(k)]  \nonumber \\
 & & \times \bigg[ \bigg\{ w\Big(-(q-\frac{P}{2})^2-k^2-l^2\Big) 
 - w\Big(-q^2-\frac{3}{4}P^2-k^2-l^2+Pl\Big) \bigg\} \nonumber \\
 & & {} \quad \times \epsilon^{ff_2f_3}\epsilon^{gg_2g_3} 
 {\rm tr}^{\rm (DC)}[(S_F^Ri\frac{\lambda^\alpha}{2})_{g_3f_3}(l)]
 \nonumber \\
 & & {} + \bigg\{ w\Big(-(q+\frac{P}{2})^2-k^2-l^2\Big) - w\Big(-q^2
 -\frac{3}{4}P^2-k^2-l^2-Pl\Big) \bigg\} \nonumber \\
 & & {} \quad \times \epsilon^{ff_2f_3}\epsilon^{gg_2g_3} 
 {\rm tr}^{\rm (DC)}[(i\frac{\lambda^\alpha}{2}S_F^R)_{g_3f_3}(l)] \bigg].
  \label{AEQ:H110315:4}
\end{eqnarray}
The term given in terms of $E^\alpha(q;P)$ and $F^\alpha(q;P)$ represents 
the correction to the Noether axialvector current due to the momentum 
dependencies of the effective lagrangian.
 For the pion, we can choose 
the flavor structure matrix $\lambda^\alpha$ of the Noether current 
in the above formula to match that of the BS amplitude. 
Then we obtain the decay constant $f_\pi$ in the usual sense.
On the other hand, for the $\eta$ or $\eta'$ mesons, 
the flavor structure of the BS amplitude depends on the relative and
total momenta in general. Therefore we can not fix $\lambda^\alpha$
from the flavor structure of the BS amplitude. 
Instead we only consider the decay constants associated with the 
octet ($\alpha=8$) and singlet ($\alpha=0$) axialvector currents 
for the $\eta$ or $\eta'$mesons, i.e., $f_8^{\eta}$, $f_0^{\eta}$,
$f_8^{\eta'}$ and $f_0^{\eta'}$.  
The fact that the flavor structure of the $\eta$-$\eta'$ meson BS amplitudes
depend on the relative and total momenta means that one cannot define the 
$\eta$-$\eta'$ mixing angle.  It can be defined only in the limit of 
neglecting these momentum dependences. 
\section{Nambu--Goldstone Solution}\label{SEC:NGS}
A remark is given here about the Nambu--Goldstone solution.
In the chiral limit, the effective action is invariant 
under the $SU_L(3)\times SU_R(3) \times U_V(1)$ transformation.
Under the dynamical breakdown of this symmetry to $SU_V(3)\times U_V(1)$,
we expect eight Nambu--Goldstone (NG) solutions.
This can be proved by the same procedure as in Ref.\cite{NYNOT1}.
However, we show here the existence of these NG solutions directly
from the SD equation $(\ref{AEQ:H091214:2})$ and 
the BS equation $(\ref{AEQ:H110209:3})$.
Multiplying the $\gamma_5 \lambda^\alpha/2$ from left and $\gamma_5
\lambda^\alpha/2$ from right to Eq.$(\ref{AEQ:H091214:2})$,
we obtain
\begin{eqnarray}
\lefteqn{
 \{ i\gamma_5\frac{\lambda^\alpha}{2}, S_F^{R-1}(q) \}_{nm}
 - \{ i\gamma_5\frac{\lambda^\alpha}{2} , S_0^{R-1} (q) \}_{nm}
} \nonumber \\
 & = & -\frac{iC_F}{f(-q^2)} \int_p \frac{1}{f(-p^2)} 
 \bar{g}^2(q_E^2,p_E^2)
 iD^{\mu\nu}(p-q) \left( \gamma_\mu \{ S_F^{R}(p),
 i\gamma_5\frac{\lambda^\alpha}{2} \} \gamma_\nu \right)_{nm} 
 \nonumber \\
 & & {} +G_D \delta_{ji} \delta_{ba} \left(
  \frac{(\lambda^\alpha)_{gg'}}{2}\epsilon^{g'f_1f_2} 
  \epsilon^{fg_1g_2}+ \epsilon^{gf_1f_2}\epsilon^{f'g_1g_2} 
  \frac{(\lambda^\alpha)_{f'f}}{2}\right) \nonumber \\
  & & {} \times \int_{p,k} \frac{1}{f(-p^2)f(-k^2)}w(-q^2-p^2-k^2)
  {\rm tr}^{\rm (DC)}[S^R_{Fg_1f_1}(p)] 
  {\rm tr}^{\rm (DC)}[S^R_{Fg_2f_2}(k)]. 
  \label{AEQ:H110210:1}
\end{eqnarray}
In the chiral limit the second term of the left hand side vanishes. 
Furthermore if $\lambda^\alpha$ is an octet matrix i.e. 
${\rm tr}[\lambda^\alpha]=0$, then a relation
\begin{eqnarray}
\lefteqn{
 \left( \frac{(\lambda^\alpha)_{gg'}}{2}
 \epsilon^{g'f_1f_2}\epsilon^{fg_1g_2}  
 +  \epsilon^{gf_1f_2}\epsilon^{f'g_1g_2} 
 \frac{(\lambda^\alpha)_{f'f}}{2}
 \right) \delta_{g_1f_1} \delta_{g_2f_2} } \nonumber \\
 & = & -4 \epsilon^{gf_1f_2} \epsilon^{fg_1g_2} \delta_{g_1f_1} 
 \frac{(\lambda^\alpha)_{g_2f_2}}{2} \label{AEQ:H110210:2}
\end{eqnarray}
leads us to 
\begin{eqnarray}
\lefteqn{
 \{ i\gamma_5\frac{\lambda^\alpha}{2}, S_F^{R-1}(q) \}_{nm}
} \nonumber \\
 & = & -\frac{iC_F}{f(-q^2)} \int_p \frac{1}{f(-p^2)} 
 \bar{g}^2(q_E^2,p_E^2)
 iD^{\mu\nu}(p-q) \left( \gamma_\mu \{ S_F^{R}(p),
 i\gamma_5\frac{\lambda^\alpha}{2} \} \gamma_\nu \right)_{nm} 
 \nonumber \\
 & & {} + 2i G_D \delta_{ji} \delta_{ba} 
  \epsilon^{gf_1f_2} \epsilon^{fg_1g_2}
  \int_{p,k} \frac{1}{f(-p^2)f(-k^2)}w(-q^2-p^2-k^2) \nonumber \\
  & & {} \times
  {\rm tr}^{\rm (DC)}[S^{R}_{Fg_1f_1}(p)] 
  {\rm tr}^{\rm (DC)}[\gamma_5 \{i\gamma_5\frac{\lambda^\alpha}{2},
  S^{R}_{F}(k)\}_{g_2f_2}]. 
  \label{AEQ:H110210:3}
\end{eqnarray}
Comparing this with Eq.$(\ref{AEQ:H110209:3})$, we obtain the 
Nambu--Goldstone solution for $\alpha = 1,\dots, 8$
\begin{equation}
 \chi^R_{nm}(q;P=0) = N \{i\gamma_5\frac{\lambda^\alpha}{2} 
 ,f(-q^2) S^R_F(q) \}
 \label{AEQ:H110210:4}
\end{equation}
where $N$ is a normalization constant.
It can be also shown that $N$ equals to $1/f_B$ where $f_B$ is 
a decay constant defined in the previous section in the framework of 
the present ladder like approximation. It should also be noted that  
Eq.$(\ref{AEQ:H110210:4})$ does not hold for a singlet 
$\lambda^{\alpha=0}$.

\section{Anomalous PCAC Relation}\label{SEC:PCAC}
 The matrix element of the PCAC relation $(\ref{AEQ:H100126:3})$
between a meson state $\langle \Kmbf{P} |\,$ and the vacuum $|0\rangle$
in the ladder like approximation becomes 
\begin{equation}
 -f_B^\alpha M_B^2 = 2[m_0{\cal E}_B^\alpha] + \delta^{\alpha 0} {\cal A}_B
 \label{AEQ:H110315:6}
\end{equation}
where $f_B^\alpha$ is the decay 
constant $f_\pi,f_8^\eta,f_0^\eta,f_8^{\eta'}$ or $f_0^{\eta'}$.
$[m_0 {\cal E}_B^\alpha]$ and ${\cal A}_B$ are defined by
\begin{equation}
 [m_0 {\cal E}_B^\alpha] := \lim_{P\to P_B} i\int_q 
\frac{f(-q_-^2)+f(-q_+^2)}{2f(-q_-^2)f(-q_+^2)} {\rm tr}[
 \overline{\chi}^R(q;P)m_0 \gamma_5\frac{\lambda^\alpha}{2}],
 \label{AEQ:H110322:1}
\end{equation}
\begin{equation}
 {\cal A}_B := \lim_{P\to P_B} \int_q \frac{1}{f(-q_+^2)f(-q_-^2)}
 {\rm tr}[\overline{\chi}^R(q;P_B) A(q;P) ],
 \label{AEQ:H110315:7}
\end{equation}
\begin{eqnarray}
\lefteqn{
 A_{mn}(q;P) := -3 G_D(\gamma_5)_{ab}\delta_{ij}
  \int_{k,l} \frac{1}{f(-k^2)f(-l^2)} 
 {\rm tr}^{\rm (DC)}[S^R_{Fg_2f_2}(k)] }\nonumber \\
 & & \times \bigg[ w\Big( -(q-\frac{P}{2})^2 -k^2 - l^2\Big) 
 \epsilon^{ff_2f_3}\epsilon^{gg_2g_3}{\rm tr}^{\rm (DC)}
 [(S^R_Fi\frac{\lambda^\alpha}{2})_{g_3f_3}(l)] \nonumber \\
 & & {} + w\Big( -(q+\frac{P}{2})^2 - k^2-l^2\Big) \epsilon^{ff_2f_3}
 \epsilon^{gg_2g_3}{\rm tr}^{\rm (DC)} [(i\frac{\lambda^\alpha}{2}
  S^R_F)_{g_3f_3}(l)] \bigg]
  \label{AEQ:H110315:8}
\end{eqnarray}
respectively.
This relation is obtained systematically using 
the method of Sec.3 in Ref.\cite{NYNOT1}. 
Of course, we can also obtain this relation (\ref{AEQ:H110315:6})
directly from the SD equation $(\ref{AEQ:H091214:2})$ 
and the BS equation $(\ref{AEQ:H110209:3})$.
If we employ the BS amplitude in the chiral limit like 
Eq.(\ref{AEQ:H110210:4}), it holds that 
\begin{equation}
 [m_0 {\cal E}_B^\alpha] = m_q \langle \bar q q \rangle_0 / f_B
 = m_{qR} \langle \bar q q \rangle_R / f_B.
 \label{AEQ:H110714:1}
\end{equation}
Eq.(\ref{AEQ:H110315:6}) with Eq.(\ref{AEQ:H110714:1}) leads us to 
the Gell-Mann, Oakes and Renner mass formula 
\begin{equation}
 M_B^2 f_B^2 \simeq -2m_q \langle \bar q q \rangle
 \quad \mbox{for} \quad \alpha\ne 0 .
 \label{AEQ:H110714:2}
\end{equation}
\par
For later use, we define the ratio
\begin{equation}
 {\cal R}_\alpha(P_E^2) = \frac{ f_B^\alpha(P_E^2) P_E^2 }
 {2[m_0 {\cal E}^\alpha_B(P_E^2)] 
 + \delta^{\alpha 0}{\cal A}_B(P_E^2) },
 \label{AEQ:H110323:3}
\end{equation}%
which is to be unity at the on-mass-shell point of the Bethe-Salpeter 
solution.  This condition is useful in checking the numerical extrapolation 
procedure.
\section{Numerical Results}\label{SEC:NR}
 In the present model, there are seven input parameters.
 Five of them are the parameters of the improved ladder 
approximation (ILA) model of QCD: 
the current quark mass $m_{qR}$ for the up and down quarks, 
the scale parameter of QCD 
$\Lambda_{\rm QCD}$, the infrared cut-off $t_{\rm IF}$ for the running 
coupling constant, the smoothness parameter $t_0$ and the ultraviolet
cut-off $\Lambda_{\rm UV}$. 
We take the value of  $t_0$ from the result of Ref.\cite{KG1}, namely, 
$t_0 = -3$. As explained there this smoothness parameter $t_0$ is introduced 
just for the stability of the numerical calculations and has no physical 
meanings. 
We take $\Lambda_{\rm UV} = 100$ [GeV] because the physical observables 
depend on it rather weakly after the renormalization as far as we 
use a reasonably large value of $\Lambda_{\rm UV}$.  
The renormalization point $\mu$ is taken as $\mu = 2$ [GeV].
The infrared cut-off $t_{\rm IF}$ controls the strength of the running 
coupling constant in the low $q^2_E$ region. Therefore its value is 
directly related to the size of the dynamical chiral symmetry breaking.
We take $t_{\rm IF} = - 0.5$.  In the case of no FMI, $t_{\rm IF} = - 0.5$
gives $- \langle \bar \psi \psi \rangle^{1/3}_R = 259$ [MeV] with 
$\Lambda_{\rm QCD} = 600$ [MeV] in the chiral limit.
\par
    We choose $\Lambda_{\rm QCD}=600$ [MeV]. 
Although this value is somewhat larger than the ``standard'' value 
$\Lambda_{\rm QCD}=100 \sim 300$ [MeV], 
it is necessary for the rainbow gluon self energy to generate the 
dynamical chiral symmetry breaking strongly\cite{KG1,NYNOT3}.
Of course if the dynamical chiral symmetry breaking is caused by
the flavor mixing interaction mainly, we may choose a smaller 
$\Lambda_{\rm QCD}$.
But in such a situation, a careful analysis is necessary.
Since the phase transition is of the first order, there may exist 
multiple SD solutions.
We will report such results elsewhere \cite{NTO}. 
Therefore in this paper we concentrate only on the case that 
the chiral symmetry breaking is generated mainly by the gluon 
exchange interaction.
\par
    We have two new parameters associated with the flavor mixing 
interaction (FMI), i.e., $G_D$ and $\kappa$.
Instead of $G_D$, we use the parameter $I_G$ defined by
\begin{equation}
 G_D \, \mbox{[GeV$^{-5}$]} \,=: -(I_G\, \mbox{[GeV$^{-1}$]} )^5.
  \label{AEQ:H110304:1}
\end{equation}
This parameter is chosen freely so that we study the effects of the
$U_A(1)$ anomaly on the $\eta$-$\eta'$ system.
The $\kappa$ parameter is taken as 
\begin{equation}
 \kappa = 0.7 \,\mbox{[GeV$^{-2}$]}.   \label{AEQ:H110304:2}
\end{equation} 
This value corresponds to the form factor of the instanton of the average 
size $\rho$, about $1/3$ [fm]. The instanton form factor,
\begin{equation}
 \frac{1}{x_E^2+\rho^2} \propto 1 - \frac{x_E^2}{\rho^2} + \cdots
 \label{AEQ:H110305:1}
\end{equation}
can be identified with the Fourier transformation of the weight function
\begin{equation}
 {\rm F.T.}\, w(q_E^2) = C \exp\left( -\frac{x_E^2}{4\kappa} \right) 
 \propto 1 - \frac{x_E^2}{4\kappa} + \cdots
 \label{AEQ:H110305:2}
\end{equation}
with
\begin{equation}
 4\kappa = \rho^2.
 \label{AEQ:H110305:3}
\end{equation}
\par
    The values of the model parameters we use throughout this article are 
$\Lambda_{\rm UV} = 100$ [GeV], $\Lambda_{\rm QCD} = 600$ [MeV],
$t_0 = -3$, $t_{\rm IF} = -0.5$, $\mu = 2$ [GeV] and $\kappa = 0.7$ 
[GeV$^{-2}$].
\par
\begin{figure}[tbp]
  \centerline{ \epsfxsize=10cm \epsfbox{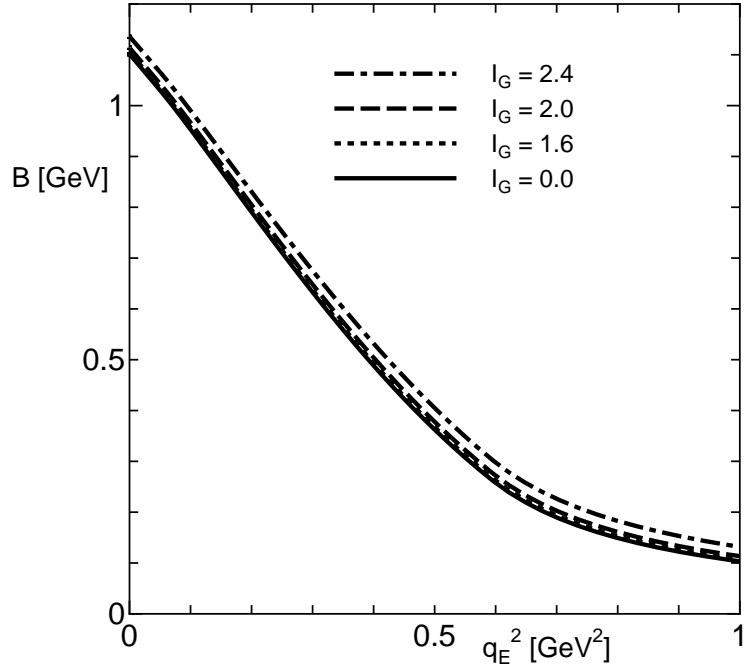} }
  \caption{$q_E^2$ dependences of the solutions of the SD equation 
  in the $SU(3)$ chiral limit with $I_G = 0.0, 1.6, 2.0, 2.4$ [GeV$^{-1}$]. }
  \label{FIG:H110210:1}
\end{figure}
\begin{figure}[tbp]
  \centerline{ \epsfxsize=10cm \epsfbox{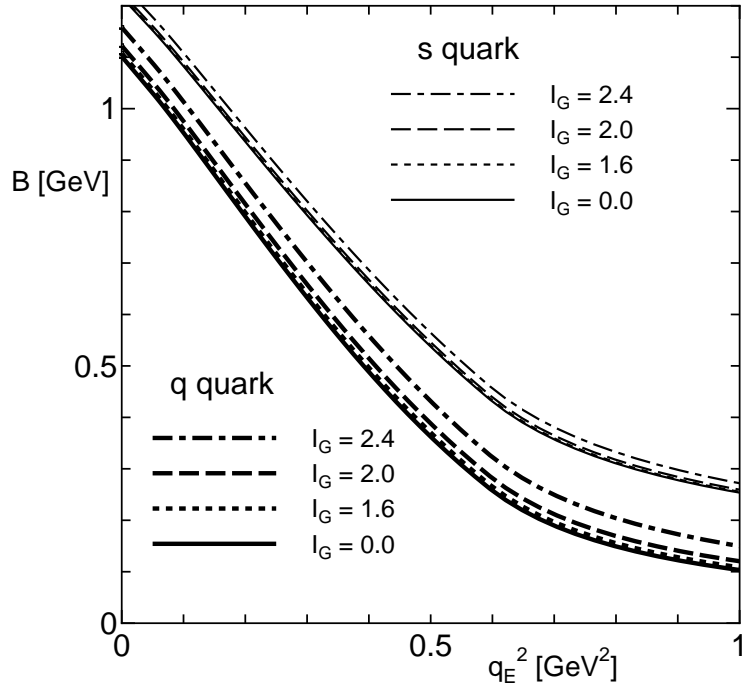} }
  \caption{$q_E^2$ dependences of the solutions of the SD equation in 
  the $SU(2)$ chiral limit with $m_{sR}=100$ [MeV] and 
  $I_G = 0.0, 1.6, 2.0, 2.4$ [GeV$^{-1}$]. }
  \label{FIG:H110210:2}
\end{figure}

    Let us now discuss the solutions of the SD equation.  
Our numerical results are shown in Table \ref{TBL:H110301:1} and
Figs.\ref{FIG:H110210:1} and \ref{FIG:H110210:2}.
As can be seen from them, the chiral symmetry breaking is induced mainly by 
the gluon exchange interaction, and the effect of FMI to the chiral quark  
condensate seems very small.
When $I_G$ increases from zero to 2.4, $B_q(0)$ increases only
4-6\%, and $\langle \bar qq \rangle_R$ changes by about 10 to 20 \%.  
One may wonder whether the range of variation is too small for its effects 
to be seen, but, as we will see later, this $I_G$ gives a large mass to 
$\eta$ and $\eta'$.
Since the perturbative quark mass contribution to the quark condensate is 
subtracted in our definition, the absolute value of 
$\langle \bar ss \rangle_R$ is smaller than that of $\langle \bar qq \rangle_R$
($q = u,d$).  Our results in $m_{qR} = 5$ [MeV] and $m_{sR} = 100$ [MeV] are 
$\langle \bar ss \rangle_R / \langle \bar qq \rangle_R = 0.85$ and 0.63 for 
$I_G = 0$ and 2.4 [GeV$^{-1}$] respectively.
These are in reasonable agreement with the QCD sum rule results: 
$\langle \bar ss \rangle_R / \langle \bar uu \rangle_R = 0.8 \pm 0.1$ 
\cite{RRY85} and 
$\langle \bar ss \rangle_R / \langle \bar uu \rangle_R = 0.6 \pm 0.1$
\cite{N89}.
The absolute value of the u,d-quark condensate increases as $I_G$ increases, 
while that of the s-quark condensate decreases as $I_G$ increases. 
It is an interesting feature in the present model. 
In the case of the Nambu-Jona-Lasinio (NJL) model with FMI, both condensates
increase.  We do not find out the intuitive explanation of the 
difference of this behavior between the NJL model and the present model.
\begin{table}[tbp]
\begin{center}
  \begin{tabular}{ccc|cccc} \hline \hline
    $m_{qR}$[MeV] & $m_{sR}$[MeV] & $I_G$[GeV$^{-1}$] & $B_q(0)$ & $B_s(0)$ &
    $\langle \bar q q \rangle_R$[GeV$^3$] & $\langle \bar s s \rangle_R$
    [GeV$^3$] \\ \hline  
    0 & 0 & 0.0 & $1.10$ & $1.10$ & $-(0.259)^3$ & $-(0.259)^3$ \\
    0 & 0 & 1.6 & $1.11$ & $1.11$ & $-(0.260)^3$ & $-(0.260)^3$ \\
    0 & 0 & 2.0 & $1.11$ & $1.11$ & $-(0.262)^3$ & $-(0.262)^3$ \\
    0 & 0 & 2.4 & $1.14$ & $1.14$ & $-(0.268)^3$ & $-(0.268)^3$ \\ \hline
    0 & 100 & 0.0 & $1.10$ & $1.22$ & $-(0.259)^3$ & $-(0.245)^3$ \\
    0 & 100 & 1.6 & $1.11$ & $1.22$ & $-(0.261)^3$ & $-(0.244)^3$ \\
    0 & 100 & 2.0 & $1.12$ & $1.22$ & $-(0.266)^3$ & $-(0.241)^3$ \\
    0 & 100 & 2.4 & $1.16$ & $1.24$ & $-(0.276)^3$ & $-(0.235)^3$ \\ \hline
    5 & 100 & 0.0 & $1.11$ & $1.22$ & $-(0.259)^3$ & $-(0.245)^3$ \\
    5 & 100 & 1.6 & $1.11$ & $1.22$ & $-(0.261)^3$ & $-(0.244)^3$ \\
    5 & 100 & 2.0 & $1.13$ & $1.23$ & $-(0.266)^3$ & $-(0.241)^3$ \\
    5 & 100 & 2.4 & $1.17$ & $1.24$ & $-(0.276)^3$ & $-(0.236)^3$ \\ 
  \hline \hline	
  \end{tabular}
\end{center}
  \caption{Dependences of the solutions of the SD equations at $q_E^2 =0$ and 
  the quark condensates on the strength of FMI with three sets of 
  the quark masses, (i) $m_{qR} = m_{sR} = 0$, (ii) $m_{qR} = 0$, $m_{sR} = 100$ [MeV] and 
  (iii) $m_{qR} = 5$ [MeV], $m_{sR} = 100$ [MeV].  
  Here q represents the u and d quarks.} 
  \label{TBL:H110301:1}
\end{table}
\par
    Let us now turn to the discussion of the solutions of the BS equation.
Our numerical results for the pion are summarized in Table \ref{TBL:H110316:1}.
Here we have not performed the precise parameter fittings so as to reproduce
the observed pion mass and decay constant since solving the BS equation 
of the non-local interaction requires the rather large computer resouces.
We observe that the pion mass and decay constant are not sensitive to 
the flavor mixing interaction. 
The deviation of the slope $M_{\pi}^2 /m_{qR}$ from the slope 
$-2 \langle \bar qq \rangle_R / f_\pi^2$ derived 
from Gell-Mann-Oakes-Renner (GMOR) relation is about 16\% at $m_{qR} = 5$ 
[MeV] 
and $I_G = 2.4$ [GeV$^{-1}$].  We consider this amount of the deviation of 
the GMOR relation may come from the Euclid $\to$ Minkowski extrapolation, 
since the ratio ${\cal R}$ defined in 
Eq.(\ref{AEQ:H110323:3}) deviates from unity by 5\%, which indicates 
the size of the numerical error for $M_\pi^2$ and $f_\pi$ in the 
extrapolation procedure.
\begin{table}[tbp]
\begin{center}
  \begin{tabular}{ccc|ccccc} \hline \hline
    $m_{qR}$[MeV] & $m_{sR}$[MeV] & $I_G$[GeV$^{-1}$] &  
    $M_{\pi}$[MeV]  & $f_{\pi}$[MeV] & ${\cal R}$ \\ \hline 
    0 & 0 & $0.0$ & $0$ & $86$ & --- \\
    0 & 0 & $1.6$ & $0$ & $87$ & --- \\
    0 & 0 & $2.0$ & $0$ & $89$ & --- \\
    0 & 0 & $2.4$ & $0$ & $95$ & --- \\ \hline
    0 & 100 & $0.0$ & $0$ & $86$ & --- \\
    0 & 100 & $1.6$ & $0$ & $88$ & --- \\
    0 & 100 & $2.0$ & $0$ & $91$ & --- \\
    0 & 100 & $2.4$ & $0$ & $101$ & --- \\ \hline
    5 & 100 & $0.0$ & $159$ & $88$ & $1.06$ \\
    5 & 100 & $1.6$ & $158$ & $90$ & $1.06$ \\
    5 & 100 & $2.0$ & $157$ & $94$ & $1.05$ \\
    5 & 100 & $2.4$ & $152$ & $103$ & $1.05$ \\
  \hline \hline	
  \end{tabular}
\end{center}
  \caption{Dependences of the solutions of the pion BS equation on the 
  strength of FMI with three sets of the quark masses, (i) $m_{qR} = m_{sR} = 0$, 
  (ii) $m_{qR} = 0$, $m_{sR} = 100$ [MeV] and 
  (iii) $m_{qR} = 5$ [MeV], $m_{sR} = 100$ [MeV]. }
  \label{TBL:H110316:1}
\end{table}
\par
  The BS solutions for the $\eta$ and $\eta'$ mesons are given 
in Tables \ref{TBL:H110303:1} and \ref{TBL:H110301:2}.
Since the BS equation is homogeneous, the absolute sign
of the BS amplitudes, and therefore the decay constants, cannot 
be determined. 
We choose the sign of $f_8$ ($f_0$) to be positive for $\eta$ ($\eta'$). 
The masses of $\eta$ and $\eta'$ and their decay constants
depend strongly on the flavor mixing interaction. Especially, the $\eta'$
meson mass seems sensitive to the flavor mixing interaction.
This is in contrast to the pion result. 
$U_A(1)$ symmetry breaking gives a large effect
on the $\eta$ and $\eta'$ sector.
\par
   In order to see the effects of the flavor mixing, we introduce
the mixing angles for the $\eta$ and $\eta'$ mesons,
\begin{equation}
\frac{- f_0^\eta}{f_8^\eta} = \tan \theta_\eta , \quad 
\frac{f_8^{\eta'}}{f_0^{\eta'}} = \tan \theta_{\eta'} .
\label{AEQ:H1106014:1}
\end{equation}
The results are presented in Tables \ref{TBL:H110303:1} and 
\ref{TBL:H110301:2}.  Since the flavaor structure of the $\eta$-$\eta'$ 
meson BS amplitudes depend on the relative and total momenta, 
the above definitions of the mixing angles are the kinds of the averaged 
quantities. 
\par
    In the $SU(3)$ symmetry limit, no flavor mixing occurs and 
$\theta_\eta = \theta_{\eta'} = 0$.  On the other hand, in the broken 
$SU(3)$ case without FMI, $\eta$ and $\eta'$ are in the ideally mixed 
states, i.e., $\theta = \arctan(- \sqrt{2}) = -54.7^\circ$. 
The mixing angles $\theta_\eta$ 
and $\theta_{\eta'}$ increase as FMI becomes strong. 
In the case of $m_{qR} = 5$ [MeV], $m_{sR} = 100$ [MeV] and $I_G = 2.4$ 
[GeV$^{-1}$], the reasonable values of the $M_\eta$ and $M_{\eta'}$ 
are obtained, the calculated $M_\eta$ is 7\% smaller than the observed 
$M_\eta$ and the calculated $M_{\eta'}$ is 11\% larger than the 
observed $M_{\eta'}$. In this model parameters, the calculated mixing angle 
for the $\eta'$ meson $\theta_{\eta'}$ is more than 5/3 times of the 
calculated mixing angle for the $\eta$ meson $\theta_\eta$. 
It means that the momentum dependences of the flavor structures of 
the $\eta$ and $\eta'$ mesons are not so small and the momentum 
independent treatment of the $\eta$-$\eta'$ mixing angle is rather 
questionable.
\par
\begin{figure}[tbp]
  \centerline{ \epsfxsize=12cm \epsfbox{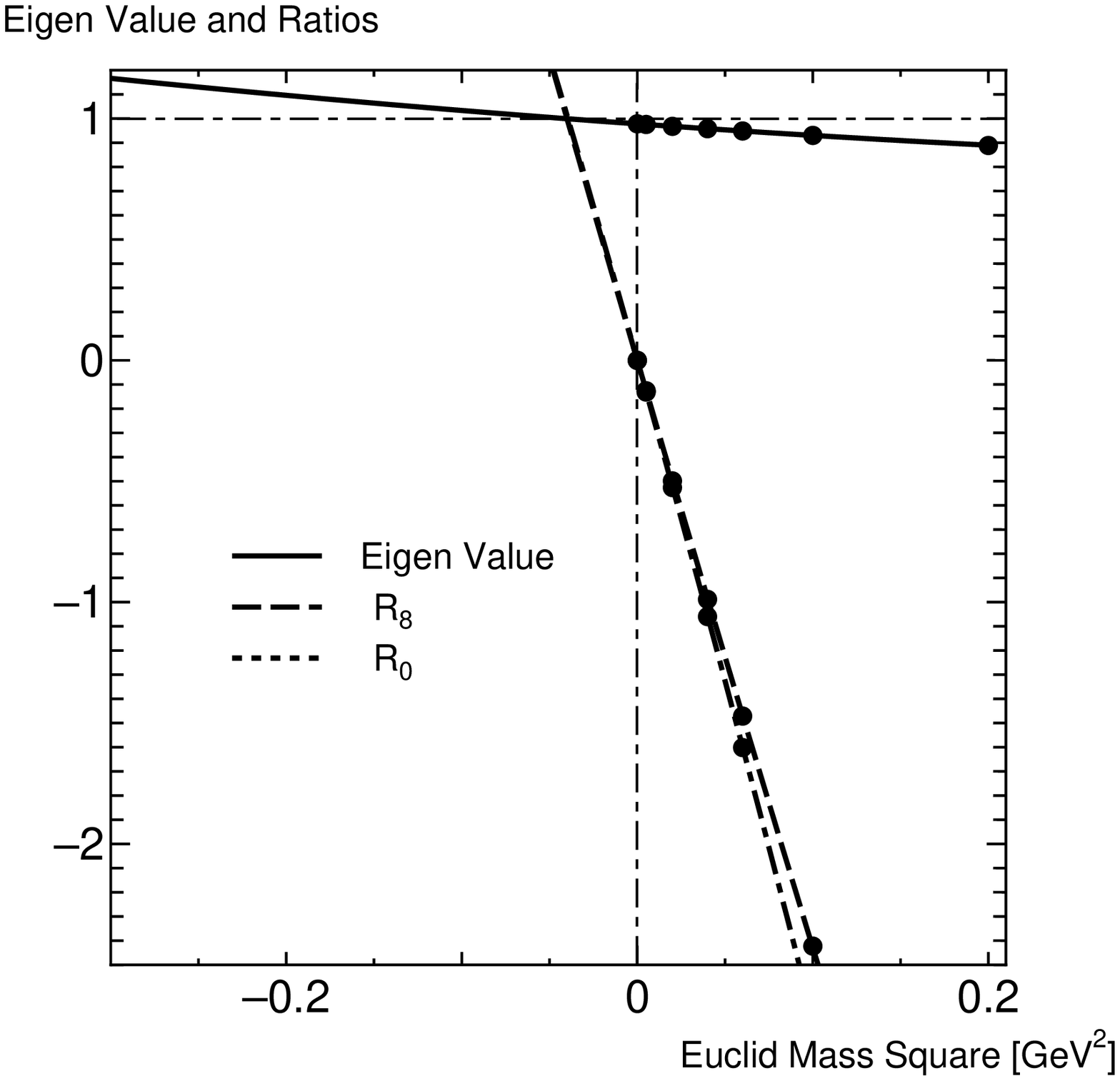} }
  \caption{Extrapolation of the eigenvalue and the ratios for the $\eta$ 
  meson with $m_{qR}=0$,\,$m_{sR}=100$ [MeV] and $I_G = 1.6$ [GeV$^{-1}$]. }
  \label{FIG:H110527:1}
\end{figure}
\begin{figure}[tbp]
  \centerline{ \epsfxsize=12cm \epsfbox{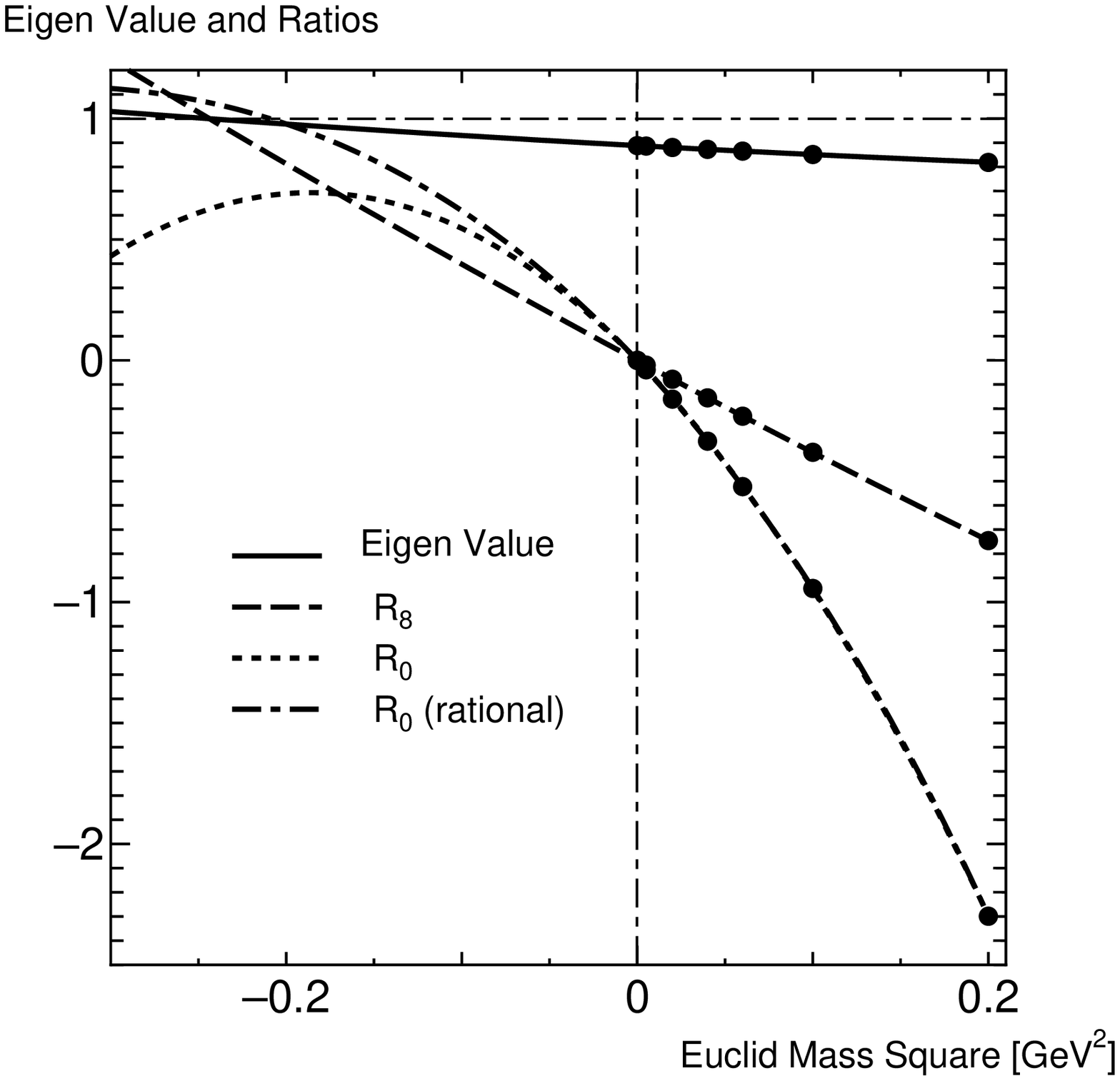} }
  \caption{Extrapolation of the eigenvalue and the ratios for the $\eta$ 
  meson with $m_{qR}=0$, $m_{sR}=100$ [MeV] and $I_G = 2.4$ [GeV$^{-1}$].}
  \label{FIG:H110304:1}
\end{figure}
\begin{figure}[tbp]
  \centerline{ \epsfxsize=12cm \epsfbox{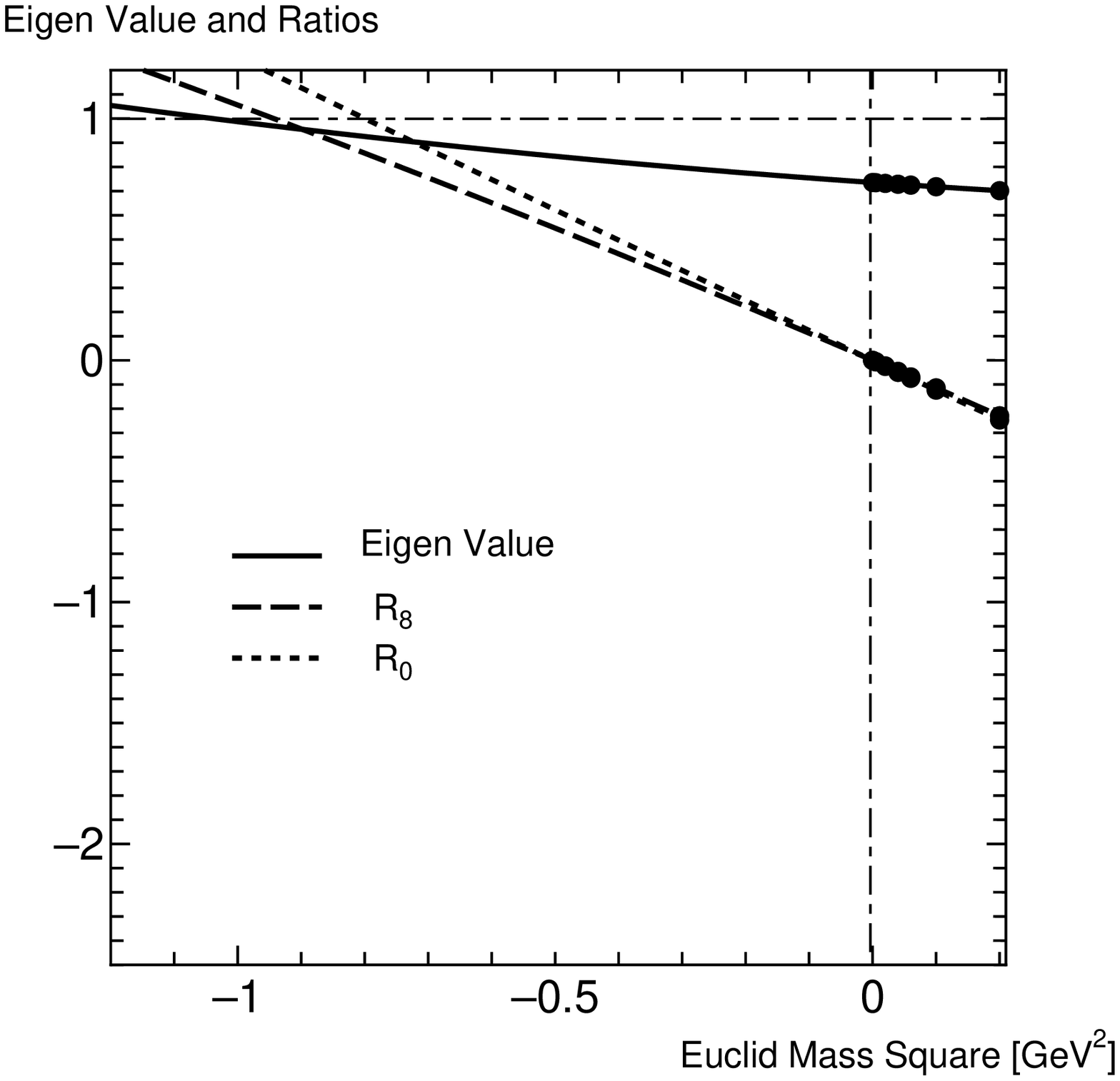} }
  \caption{Extrapolation of the eigenvalue and the ratios for the $\eta'$ 
  meson with $m_{qR}=0$, $m_{sR}=100$ [MeV] and $I_G = 2.4$ [GeV$^{-1}$].}
  \label{FIG:H110611:1}
\end{figure}

    The last column of Tables \ref{TBL:H110316:1}, \ref{TBL:H110303:1} 
and \ref{TBL:H110301:2} gives the ratio $\cal R$ in 
Eq.(\ref{AEQ:H110323:3}) for finite quark mass. 
As it should be $1$ at the on-mass-shell point identically, 
it is a good indicator of the ambiguity, or error, coming from 
the extrapolation from the Euclidean kinematics to the Minkowski 
on-mass-shell kinematics.  
Here we carry out the quadratic extrapolations of the eigenvalue $\lambda$
and the ratio $\cal R$ and the linear extrapolation of the decay constant 
$f$ from the Euclid region to the on-mass-shell point.
In the case of the weak FMI, our extapolation procedure works rather well.
The quality of the extrapolation can be seen in Fig.\ref{FIG:H110527:1}.
For heavier meson masses, ${\cal R}$ deviates from 1 
significantly. This indicates an extrapolation error.
In fact, for $M_B > 700$ [MeV] the extrapolation becomes very 
difficult in the quadratic extrapolation.
For instance, ${\cal R}_0^\eta$ for $I_G=2.4$ [GeV$^{-1}$] shown in 
Table \ref{TBL:H110303:1} largely deviates from 1. 
Fig.\ref{FIG:H110304:1} shows the extrapolation in this case,
where the lines of $\lambda$ and ${\cal R}_8^\eta$ are almost  
straight but the curve of ${\cal R}_0^\eta$ is not.
This may be a reason why ${\cal R}_0^\eta$ deviates from unity. 
We have performed the extrapolation of ${\cal R}_0^\eta$ by using a 
rational function which is shown in Fig.\ref{FIG:H110304:1} by the 
dotted line and the result is improved well.
As for the $\eta'$ meson, the extrapolation of the eigenvalue, the 
rations ${\cal R}_8^{\eta'}$ and ${\cal R}_0^{\eta'}$ are shown in 
Fig.\ref{FIG:H110611:1}. 
\par
    From Table \ref{TBL:H110301:2} one can see that in the chiral limit, 
$\eta'$ state is a pure flavor singlet state and has finite mass.  
This mass is due to FMI and it means $\eta'$ is not the Goldstone boson. 
One of the interesting question is that how much $\eta'$ loses the
the Goldstone boson nature.  
In the present range of the $U_A(1)$ breaking interaction strength,
the flavor singlet pseudoscalar meson state has mass from 194 [MeV] 
to 634 [MeV].  On the other hand the decay constant changes only less
than 8\%.  Further studies of the $\eta'$ properties such as the 
decay amplitudes are necessary in order to understand the nature of 
the $\eta'$ meson. \par
\begin{figure}[tbp]
  \centerline{ \epsfxsize=15cm \epsfbox{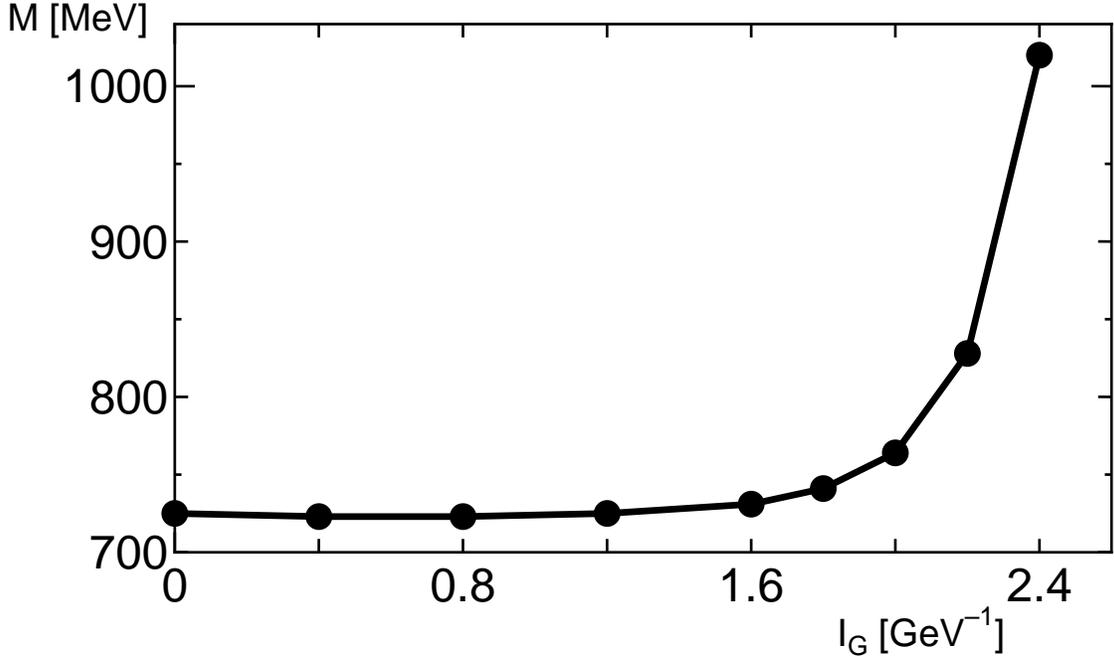} }
  \caption{$I_G$ dependence of the mass of the $\eta'$ meson with 
           $m_{qR}=0$ and $m_{sR}=100$ [MeV]. }
  \label{FIG:H110527:2}
\end{figure}

 
    We plot the $\eta'$ meson mass as a function of the $U_A(1)$ 
breaking parameter $I_G$ in Fig.\ref{FIG:H110527:2}. The effect of
the mixing of $u, d$ quark component seems to be negligible and
 the $\eta'$ mass grows rapidly from $I_G \sim 2.0$ [GeV$^{-1}$].\par  
\begin{table}[tbp]
\begin{center}
  \begin{tabular}{ccc|cccccc} \hline \hline
    $m_{qR}$ & $m_{sR}$ & $I_G$ & 
    $M_{\eta}$ & $f_8^{\eta}$ & $f_0^{\eta}$ & $\theta_\eta$ & 
    ${\cal R}_8^{\eta}$ & ${\cal R}_0^{\eta}$ \\  
    $$[MeV] & $$[MeV] & [GeV$^{-1}$] & $$[MeV] & $$[MeV] & $$[MeV] & 
    $$[deg] & &  \\ \hline 
    0 & 0 & $0.0$ & $0$ & $86$ & $0$ & 0 &--- & --- \\
    0 & 0 & $1.6$ & $0$ & $87$ & $0$ & 0 &--- & --- \\
    0 & 0 & $2.0$ & $0$ & $89$ & $0$ & 0 &--- & --- \\
    0 & 0 & $2.4$ & $0$ & $95$ & $0$ & 0 &--- & --- \\ \hline
    0 & 100 & $0.0$ & $0$ & $50$ & $70$ & $-54.7$ &--- & --- \\	
    0 & 100 & $1.6$ & $203$ & $56$ & $68$ & $-50.5$ & $1.03$  & $1.04$ \\
    0 & 100 & $2.0$ & $351$ & $71$ & $58$ & $-39.2$ &$1.00$  & $1.03$ \\
    0 & 100 & $2.4$ & $495$ & $106$ & $22$ & $-11.7$ & $1.00$  
	& $0.62$ ($1.07$) \\ 
    \hline
    5 & 100 & $0.0$ & $159$ & $51$ & $72$ & $-54.7$ & 1.05 & 1.05 \\
    5 & 100 & $1.6$ & $258$ & $57$ & $69$ & $-50.4$ & $1.02$ & $1.03$ \\
    5 & 100 & $2.0$ & $387$ & $75$ & $58$ & $-37.7$ & $1.00$ & $1.05$ \\
    5 & 100 & $2.4$ & $511$ & $113$ & $19$ & $-9.5$ & $1.01$ 
	& $0.55$ ($1.00$) \\
  \hline \hline	
  \end{tabular}
\end{center}
  \caption{Dependences of the $\eta$-meson solutions of the coupled channel BS
  equation on the strength of FMI with three sets of the quark masses, 
  (i) $m_{qR} = m_{sR} = 0$, (ii) $m_{qR} = 0$, $m_{sR} = 100$ [MeV] and 
  (iii) $m_{qR} = 5$ [MeV], $m_{sR} = 100$ [MeV]. The values in the brace are 
 obtained by the rational extrapolation.}
  \label{TBL:H110303:1}
\end{table}
\begin{table}[tbp]
\begin{center}
  \begin{tabular}{ccc|cccccc} \hline \hline
    $m_{qR}$ & $m_{sR}$ & $I_G$ & 
    $M_{\eta}$ & $f_8^{\eta}$ & $f_0^{\eta}$ & $\theta_\eta$ & 
    ${\cal R}_8^{\eta}$ & ${\cal R}_0^{\eta}$ \\  
    $$[MeV] & $$[MeV] & [GeV$^{-1}$] & $$[MeV] & $$[MeV] & $$[MeV] & 
    $$[deg] & &  \\ \hline 
    0 & 0   & $0.0$ & $0$ & $0$ & $86$ & 0 &--- & --- \\
    0 & 0   & $1.6$ & $194$ & $0$ & $87$ & 0 &--- & $1.04$ \\
    0 & 0   & $2.0$ & $350$ & $0$ & $88$ & 0 & --- & $1.02$ \\
    0 & 0   & $2.4$ & $634$ & $0$ & $94$ & 0 &--- & $1.06$ \\ \hline
    0 & 100 & $0.0$ & $723$ & $-101$ & $72$  & $-54.7$ & $1.05$ & $1.05$ \\	
    0 & 100 & $1.6$ & $731$ & $-99$  & $76$  & $-52.5$ & $1.05$ & $1.07$ \\
    0 & 100 & $2.0$ & $764$ & $-88$  & $88$  & $-45.0$ & $1.06$ & $1.10$ \\
    0 & 100 & $2.4$ & $1020$ & $-41$ & $118$ & $-19.2$ & $1.09$ 
	& $1.30$($1.32$) \\ 
    \hline
    5 & 100 & $0.0$ & $723$ & $-101$ & $72$  & $-54.7$ & $1.05$ & $1.05$ \\	
    5 & 100 & $1.6$ & $732$ & $-98$ & $77$ & $-51.8$ & $1.05$ & $1.07$ \\
    5 & 100 & $2.0$ & $777$ & $-86$ & $90$ & $-43.7$ & $1.06$ & $1.11$ \\
    5 & 100 & $2.4$ & $1060$ & $-36$ & $122$ & $-16.4$ &$1.09$ & $1.33$ \\ 
    \hline \hline
  \end{tabular}
\end{center}
  \caption{Dependences of the $\eta'$-meson solutions of the coupled channel BS
  equation on the strength of FMI with three sets of the quark masses, 
  (i) $m_{qR} = m_{sR} = 0$, (ii) $m_{qR} = 0$, $m_{sR} = 100$ [MeV] and 
  (iii) $m_{qR} = 5$ [MeV], $m_{sR} = 100$ [MeV]. 
  The value in the brace is obtained by the rational extrapolation.}
  \label{TBL:H110301:2}
\end{table}
\par
    The $\eta$-meson properties have been studied extensively in the 
three-flavor NJL model with the instanton induced $U_A(1)$ breaking 
flavor mixing interaction (FMI) in Ref.\cite{TO95} and it has been shown
that the $\eta$-meson mass, the $\eta \rightarrow \gamma \gamma$, 
$\eta \rightarrow \gamma$, $\mu^- \mu^+$ and 
$\eta \rightarrow \pi^0 \gamma \gamma$ decay widths are reproduced well with 
the rather strong FMI. In this case the contribution from FMI to the 
dynamical mass of the up and down quarks is about 44\% of that from the 
usual $U_L(3) \times U_R(3)$ invariant four-quark interaction.
In contrast with it, the contribution from FMI to the dynamical 
quark mass is very small in the present study.  To make the situation 
clear, let us compare the strength of FMI in the present case with 
that in the NJL model case.  The naive way is to compare the following 
two quantities,
\begin{equation}
\int_0^\infty dq e^{-\kappa q^2} I_G^5 , 
\end{equation}
in the present model and
\begin{equation}
\int_0^\Lambda dq \frac{16 \pi^4}{9 \Lambda^5} G_D^{\rm eff} , 
\end{equation}
in the NJL model.  In this manner $I_G = 2.4$ [GeV$^{-1}$] corresponds 
to $G_D^{\rm eff} = 0.73$, which is rather close to the value determined 
in Ref.\cite{TO95}, $G_D^{\rm eff} = 0.7$.  It means the strength of FMI 
is almost same in the both cases.  It is not clear why the contribution 
from FMI to the dynamical quark mass is so different in the two cases.
\par
    It should be noted here that as can be seen from Fig.\ref{FIG:H110209:2},  
FMI operates on the pseudoscalar meson states as the effective 
4-quark interaction reduced by contracting a quark-antiquark pair into a 
quark condensate.  Therefore the effective strength of FMI is not 
$G_D$ but $G_D \langle \bar qq \rangle$.  In our present 
model parameters, the dynamical chiral symmetry breaking is mostly driven by 
the one-gluon exchange type interaction.  If one reduces the strength of the 
one-gluon type interaction in the infrared region, the quark condensate becomes
smaller and the effective strength of FMI on the meson states becomes weaker. 
Therefore there is a possibility of taking rather strong FMI without
destroying the success of the present description of the $\eta$ and $\eta'$ 
meson masses. In the case where FMI is dominant 
in the infrared region, we expect that the chiral phase transition becomes 
the first order and there may exist multiple solutions of the SD equation. 
In such a situation, more careful analyses are required, which will be 
reported elsewhere \cite{NTO}.
\section{Summary and Conclusions}\label{SEC:SC}
   The improved ladder approximation of QCD has successfully described the 
low-energy properties of QCD \cite{KG1,KG2,NYNOT3}.  In this article, 
we have studied the $\eta$ and $\eta'$ mesons in this approach.  
It is expected that the $U_A(1)$ anomaly plays an important role in the 
$\eta$ and $\eta'$ mesons and therefore we have introduced the instanton 
induced $U_A(1)$ breaking 6-quark determinant interaction \cite{KM70,tH76}
in the improved ladder approximation model of QCD.  We have derived the 
Schwinger-Dyson (SD) equations for the light quark propagators and the 
Bethe-Salpeter (BS) equations for the pion, $\eta$ and $\eta'$ in the 
lowest order (rainbow-ladder) approximation using the 
Cornwall-Jackiw-Tomboulis (CJT) effective action formulation \cite{CJT74}.
\par
    Using the same model parameters of the running coupling constant used in 
\cite{NYNOT3}, we have obtained reasonable values of $M_\pi$, $M_\eta$, 
$M_{\eta'}$, $f_\pi$ and $\langle \bar qq \rangle_R$ with a relatively weak
flavor mixing interaction (FMI), for which the chiral symmetry breaking 
is dominantly induced by the soft-gluon exchange interaction.
It is in contrast with the Nambu-Jona-Lasinio (NJL) model results, 
where about 1/3 of the dynaminacl quark mass is due to FMI \cite{TO95}.
\par
    As far as we know, the $\eta'$ BS equation which includes the 
running coupling aspect of QCD and the effect of the $U_A(1)$ anomaly has not 
been solved so far.  In the case of the NJL model, the $\eta'$ mass has 
unphysical large imaginary part associated with the unphysical decay channel
$\eta' \to \bar qq$.   
On the other hand, the present model predicts a bound $\eta'$ 
although it may not perfectly confine quarks.  
The bound state is obtained as the quark mass function 
becomes rather large, $B(-q^2_E)\sim 1$ [GeV] at $q_E = 0$, in this 
model, while it is independent of the momentum in the NJL model.
\par
    Since the flavor structure of the $\eta$-$\eta'$ meson BS amplitudes 
depend on the relative and total momenta, 
one cannot define the $\eta$-$\eta'$ mixing angle unambiguously.  
It can be defined only in the limit of neglecting these momentum dependences
and it should be examined whether such an approximation is reasonable.
Our numerical results indicate that the momentum independent treatment 
of the flavor mixing is rather questionable.
\par
    In the chiral limit we can define the $\eta_0$ decay constant $f_0$ 
without any ambiguity. (Here $\eta_0$ means the pure flavor singlet state.)
Our numerical results show that $(|f_0| - f_\pi)/f_\pi \le 0.012$ in our 
parameter range though $M_{\eta_0}$ changes from 194 [MeV] to 634 [MeV]. 
There is no low-energy theorem for the $\eta_0$ decay constant $f_0$ 
since the $U_A(1)$ symmetry is explicitly broken by the anomaly.  Therefore
the present result of $f_0$ should contain the information of the  
low-energy dynamics of QCD. 
\par
    The present result is our first step towards quantitative 
understanding of the 
flavor mixing interaction.  With the BS solutions in hand, we may  
calculate static properties and decay amplitudes of $\eta$ and $\eta'$.
Our model is regarded as a low energy effective theory, which is 
consistent with chiral symmetry, its spontaneous breakdown and the 
$U_{A}(1)$ anomaly.  It should be stressed that the approximation 
used in solving and renormalizing the amplitudes also respect these 
symmetry properties.
Thus our approach is suitable for further studies of the $\eta$ and 
$\eta'$ systems, which are desirable in order to clarify the role 
of the $U_A(1)$ anomaly in the low-energy QCD.
%
%
%
%
%
\section*{Acknowledgment}
K. N. acknowledges the post doctoral fellowship by RIKEN, and Profs.
M. Ishihara and K. Yazaki for their encouragement.
This work is supported in part by the Grant-in-Aid for Scientific Research
(C)(2) 08640356 and (C)(2)11640261 of the Ministry of Education, Science,
Sports and Culture of Japan.

\end{document}